\documentclass[twocolumn,pre,showpacs,amsmath,amssymb,floatfix]{revtex4}
\usepackage{graphicx}
\usepackage{dcolumn}
\usepackage{bm}

\begin{document}

\title{Critical behavior of the contact process in annealed scale-free networks}

\author{Jae Dong Noh}
\affiliation{Department of Physics, University of Seoul ,
  Seoul 130-743, Korea}
\author{Hyunggyu Park}
\affiliation{School of Physics, Korea Institute for Advanced Study, Seoul,
130-722, Korea}

\date{\today}
\begin{abstract}
Critical behavior of the contact process is studied in annealed scale-free networks by mapping
it on the random walk problem. We obtain the analytic results for the
critical scaling, using the event-driven dynamics approach. These results
are confirmed by numerical simulations. The disorder fluctuation induced by
the sampling disorder in annealed networks is also explored. Finally, we discuss over a possible
discrepancy of the finite-size-scaling theory in annealed and quenched networks
in spirit of the droplet size scale and the linking disorder fluctuation.

\end{abstract}
\maketitle

\section{Introduction}
Critical phenomena in complex networks have been attracting a lot of
interest~\cite{Dorogovtsev07}.
Complex networks are characterized by a so-called small-world
property~\cite{WS98}. The number of neighbors of a node increases
exponentially with the distance from it.
For this property, it is believed that critical phenomena in complex
networks belong to the mean field universality class. Nevertheless, structural
heterogeneity leads to rich behaviors. For example, in scale-free~(SF) networks
having a power-law degree distribution $P(k)\sim k^{-\gamma}$~\cite{BA99}, mean-field
critical exponents may vary with the degree exponent $\gamma$~\cite{Dorogovtsev07}.

Recent studies raise an important issue on the finite-size-scaling~(FSS)
theory in complex networks~\cite{Hong07,Castellano08}. A scale-free
network with $N$ nodes has a maximum cutoff $k_{\max}$ in degree. In most
cases without any constraint, the cutoff scales as $k_{\max} \sim
N^{1/(\gamma-1)}$,
which is determined by the condition $\sum_{k>k_{\max}}P(k) = 1/N$.
This is called the natural cutoff. One may impose a forced
cutoff
\begin{equation}\label{k_max}
k_{\max} = N^{1/\omega}
\end{equation}
with the cutoff exponent $\omega > \gamma-1$.
Taking the thermodynamic limit, one should
take the limit $N\rightarrow \infty$ and $k_{\max}\rightarrow \infty$
simultaneously. This may give rise to an intricate finite-size effect~\cite{Castellano08}.

Hong, Ha, and Park~\cite{Hong07} developed a FSS theory based on the
single-parameter scaling hypothesis. Their theory predicts the values of the
FSS exponents in the Ising model (including more general
equilibrium $\phi^n$ theory)  and the contact process~(CP), respectively.
The CP is a reaction-diffusion model describing an epidemic spreading, which exhibits
a prototype nonequilibrium phase transition from an inactive phase into an active phase~\cite{Hin00}.
It has been suggested that the FSS exponents depend only on the degree exponent $\gamma$,
regardless of the cutoff if it is not strong enough $(\omega<\gamma)$. Note that this condition
includes networks with the natural cutoff $(\omega=\gamma-1)$ as well as networks with a weak forced cutoff
$(\gamma-1<\omega<\gamma)$.
These results were confirmed numerically in
the static model~\cite{Goh01} having the natural
cutoff and the uncorrelated configuration
model~(UCM)~\cite{Catanzaro05}.

Castellano and Pastor-Satorras~\cite{Castellano08}
considered the CP in the so-called random neighbor (annealed) network.
Links are not fixed but fluctuate in this annealed network. At each time
step, neighbors of a node are chosen independently and randomly according to
the degree distribution. It contrasts with a network where links are
fixed permanently in time once they are formed.
In order to stress the distinction, the former network will be
referred to as an {\em annealed} network, while the latter network
as a {\em quenched} network.
From the analysis of the survival probability at the critical point,
they found that the dynamic exponent characterizing the relaxation time
depends not only on $\gamma$ but also on $\omega$ when $\omega>\gamma-1$
(all networks with a forced cutoff) and $2<\gamma<3$.
In particular, it has been shown that there are two different characteristic time
(and also the order parameter) scales which make a single-parameter scaling impossible.
From the relaxation time scaling, the order parameter in the quasi-steady state also scales with $N$ with an
exponent depending on both $\gamma$ and $\omega$.

At a glance, the results of Refs.~\cite{Hong07} and \cite{Castellano08} seem incompatible
(single-parameter  versus two-parameter scaling and cutoff-independent  versus
cutoff-dependent scaling) when $\gamma-1<\omega<\gamma$ (weak forced cutoff) and $2<\gamma<3$ (highly heterogeneous regime). But it is not true. The FSS theory of Ref.~\cite{Hong07} concerns
a quenched scale-free network, while that of Ref.~\cite{Castellano08}
concerns an annealed scale-free network. Quenched disorder in linking topology generates
local correlations through quenched links between nodes, which are responsible
for the shift of the phase transition point and its disorder fluctuations. Therefore,
one may not rule out a possibility that the disorder fluctuations near the phase transition point may wipe away or at least significantly alter the cutoff-dependent scaling regime, see Sec.~VI.

In this paper, we present a full FSS theory
governing the critical and off-critical scaling behaviors of the CP in
annealed networks. In Sec.~II, an annealed network is introduced without any
sampling disorder and the heterogeneous mean field theory is briefly reviewed for the CP.
The critical dynamics is analyzed in Sec.~III, while the off-critical scaling is investigated in Sec.~IV. In Sec.~V, we discuss the effect of sampling disorder in annealed networks and its
self-averaging property. Finally, we summarize our results along with a brief discussion on the effect of linking disorder in quenched networks.

\section{CP in annealed networks}
We consider the annealed scale-free networks with the degree distribution
$P(k)=a k^{-\gamma}$ for $k_{\min}\leq k\leq k_{\max}$ with a normalization
constant $a$ and $P(k)=0$ elsewhere. The maximum degree $k_{\max}$ scales with network size $N$ as
in Eq.~(\ref{k_max}) and the minimum degree $k_{\min}$ is an $O(1)$ constant.
Since neighbors of each node need not be specified, an annealed network
is realized by choosing a degree sequence $\{k_1,\cdots, k_N\}$ only.

There are two different ways in choosing the degree sequence.
One may assign degree $k$ to $N_k$ nodes deterministically
in such a way that $\sum_{k'\geq k} N_{k'} = \mbox{int}[N \sum_{k'\geq k}P(k')]$ for all
$k$ in the decreasing order starting from $k_{\max}$, where $\mbox{int}[x]$ is the integer part of $x$.
One may easily show that the maximum degree realized using this assignment algorithm
is the same order in $N$ of a given $k_{\max}$ when $\omega\ge\gamma-1$.
Or, one may draw probabilistically $N$ values of $k$ in accordance with the probability
distribution $P(k)$. The probabilistic method yields an ensemble of
different samples, which makes an ensemble average necessary.
We mainly consider the annealed network realized by the deterministic
method. Sample-to-sample
fluctuations in the ensemble generated by the probabilistic method will be
discussed in Sec.~V.

The CP on the annealed SF network is defined as follows. Each node is
either occupied by a particle or empty.
A particle on a node is annihilated with
probability $p$ or branches one offspring to its {\em neighbor}, if empty,
with probability $(1-p)$. At each time step, a neighbor of a node is selected
among all other nodes with probabilities proportional to their degree. Since a
node is coupled only probabilistically with all other nodes, the mean field theory becomes exact in the annealed network.

Let $n(t)$ be the number of particles at time $t$.
Following Ref.~\cite{Castellano08} in a quasistatic approximation for large $t$, it increases by 1 with probability
\begin{equation}\label{eq:w+}
w_+ = p \lambda
\rho \sum_k \frac{  k P(k) }{\langle k \rangle} \frac{1}{1+
\lambda \rho k/ \langle k \rangle}  ,
\end{equation}
or decreases by 1 with probability
\begin{equation}\label{eq:w-}
w_- = p \rho
\end{equation}
after a time step $\Delta t = 1/N$. Here $\rho = n/N$ is the particle
density and $\lambda = (1-p)/p$ with the mean degree $\langle k \rangle$.

The transition probability $w_+$ contains a nontrivial $\rho$-dependence.
When the thermodynamic limit is taken first~\cite{Hong07} or
the density is high~($\rho \gg 1/k_{\max}$) in finite
networks~\cite{Castellano08}, one may arrive at a singular expansion
\begin{equation}\label{w+high}
w_+ / p = \lambda \rho - c \rho^{\theta-1} + \cdots
\end{equation}
with a constant $c$ and
\begin{equation}
\theta = \min\{\gamma,3\} .
\end{equation}
When the density is low~($\rho \ll 1/k_{\max}$) in finite networks,
one can expand the denominator
in Eq.~(\ref{eq:w+}) to obtain
\begin{equation}\label{w+low}
w_+ / p  = \lambda \rho - \lambda^2 g \rho^2 + \cdots
\end{equation}
where $ g = {\langle k^2\rangle}/{\langle k \rangle^2 }$ with $\langle
k^n\rangle\equiv \sum_{k} k^n P(k)$.
Note that $g$ is an $O(1)$ constant for $\gamma>3$, while it scales
as $g \sim k_{\max}^{3-\gamma} \sim N^{(3-\gamma)/\omega}$
for $2<\gamma<3$. The scaling behavior can be rewritten as
\begin{equation}\label{g_s}
g \sim k_{\max}^{3-\theta} \sim N^{(3-\theta)/\omega} ,
\end{equation}
for general $\gamma (\neq 3)$ and $\omega\ge \gamma-1$. At $\gamma=3$, $g\sim \log N$.

As the stochastic fluctuation $(\Delta \rho)/\rho$ (multiplicative diffusive noise) becomes negligible in the $N\rightarrow \infty$ limit, one can write the rate equation for the  average particle density in the continuum limit as
\begin{equation}\label{rate_eq}
\frac{d\rho}{dt} = w_+ - w_- .
\end{equation}
It is clear that the system undergoes an
absorbing phase transition at $p=p_c = 1/2$~($\lambda_c = 1$) at all values
of $\gamma>2$ in the thermodynamic limit.
The particle density near the critical point scales
as $\rho \sim (\lambda-\lambda_c)^\beta$
with the order parameter exponent $\beta =
{1}/(\theta-2)$~\cite{Hong07}. At $\gamma=3$, an additional logarithmic correction appears
as $\rho\sim (\lambda-\lambda_c)/|\log (\lambda-\lambda_c)|$.

\section{Critical dynamics}\label{CD}
We consider the CP at the critical point ($p=1/2$ or $\lambda=1$).
One may regard the particle number $n$ $(0\le n\le N)$ as a
coordinate of an one-dimensional random walker~\cite{Castellano08}.
At each time step $\Delta t = 1/N$,
the walker jumps to the right with probability $w_+$ or to the left with
probability $w_-$, or does not move with probability $1-(w_+ + w_-)$.
The walker is bounded by an absorbing wall at $n=0$ and
a reflecting wall at $n=N$. Reaching the absorbing wall, it will be trapped
there forever.

It turns out that an event-driven dynamics is useful.
In this dynamics, the walker always makes a jump at each time step
$\Delta \tau=1$ to the right or left with probabilities
\begin{equation}
\tilde{w}_{\pm} = \frac{w_{\pm}}{w_+ + w_-} .
\end{equation}
This is equivalent to the original problem if one rescales the time
with the relation
\begin{equation}\label{t_tau}
dt = \frac{1}{N}\frac{d\tau}{w_+ + w_-} .
\end{equation}

\subsection{Defect dynamics}

It is interesting to study how particles spread starting from a localized seed.
Dynamics initiated from a single particle is called the defect
dynamics~\cite{HKPP98,Hin00}. So its initial condition is $n(0)=n_0=1$.

Quantities of interest are the survival probability $P_{s}(t)$, the
probability that the system is still active at time $t$,
and $n_{s}(t)$, the number of particles averaged over
surviving samples. At the critical point, they exhibit power-law scalings
\begin{equation}
P_s(t) \sim t^{-\delta} \ \ \mbox{and} \ \ n_s(t) \sim t^{\tilde\eta}
\end{equation}
for $t<t_c(N)$ with the relaxation time scaling as
\begin{equation}
t_c \sim N^{\bar z} .
\end{equation}
At $t= t_c$, the system starts to feel its finite size and
$n_s(t)$ saturates. For $t>t_c$, $P_s(t)$ decays exponentially.
The critical exponents $\delta$, ${\tilde \eta}$, and ${\bar z}$ are universal.
Note that ${\tilde\eta}=\delta+\eta$ where $\eta$ is the particle number growing exponent for all samples.

Initially, $\rho_0=n_0/N$ is so small (much smaller than $1/k_{\max}$) that one can always use the
expansion in Eq.~(\ref{w+low}) for $w_+$.
We will confirm that this is valid throughout the defect dynamics.
Then, in the event-driven dynamics, the jumping probability
for the walker at site $n$ is given by
$$ \tilde{w}_+ = \frac{1 - g \rho}{2  - g\rho} \ \
\mbox{and} \ \ \tilde{w}_- = \frac{1}{2  - g\rho}
$$
for small $\rho$.
This shows that the walker performs biased walks toward the
absorbing wall with the drift velocity
\begin{equation}
v_{drift} \equiv \frac{dn}{d\tau}=\tilde{w}_+ - \tilde{w}_- = - \frac{g\rho}{2-g\rho} .
\end{equation}

The bias is negligible~($v_{drift}/\tilde{w}_\pm \ll 1$) during the initial stage
since $g \rho \ll 1$. Hence, for sufficiently small $\tau$, it suffices to consider the unbiased
random walk motion in the presence of the absorbing wall at $n=0$.
The effect of the absorbing wall can be taken into account by
using the image method~\cite{Fisher84}. This yields that
the surviving probability decays as
\begin{equation}\label{P_tau}
P_{s}(\tau) \simeq n_0 (\pi \tau/2)^{-1/2}
\end{equation}
and that the surviving walker spreads out diffusively as
\begin{equation}\label{n_tau}
n_s(\tau) \simeq     \sqrt{\pi\tau/2} .
\end{equation}

The diffusion velocity for the surviving walkers is given by
\begin{equation}
v_{diffuse} \equiv n_s(\tau)/\tau
\simeq \sqrt{\pi /(2 \tau)} \simeq \pi /(2 n_s) .
\end{equation}
 As $\tau$ increases,
the diffusion velocity becomes smaller while the bias becomes stronger.
The walker reaches a stationary state when the diffusion velocity and
the drift velocity are balanced.
The condition $v_{diffuse} \sim |v_{drift}|$
yields that the walker reaches the stationary state at position
\begin{equation}\label{n_sc}
n_s^\infty \sim \sqrt{N/g}
\end{equation}
and at time
\begin{equation}\label{tau_c}
\tau_c \sim {N}/{g} .
\end{equation}
This result is self-consistent with the underlying assumptions that
$\rho k_{\max}\ll 1$ and $g\rho\ll 1$.

The time scales $t$ and $\tau$ are related through
Eq.~(\ref{t_tau}). Using Eqs.~(\ref{n_tau}) and (\ref{t_tau}),
one finds that
\begin{equation}
t \simeq \int^\tau \frac{d\tau'}{ n(\tau')} \sim \sqrt{\tau} .
\end{equation}
Therefore we conclude that
\begin{eqnarray}
&&P_{s}(t) \sim n_0 t^{-1},\label{defect_P}\\
&&n_s(t) \sim    t ,\label{defect_ns}\\
&&t_c \sim \sqrt{{N}/{g}} , \label{defect_tc}
\end{eqnarray}
which leads to
\begin{eqnarray}
\delta &=& 1 ,\label{delta}\\
\tilde{\eta} &=& 1 \quad(\eta=0),\label{eta}\\
{\bar z} &=& (1-(3-\theta)/\omega)/2 . \label{z_d}
\end{eqnarray}
The result for $\delta$ and ${\bar z}$ coincides with that of
Ref.~\cite{Castellano08}. At $\gamma=3$, $t_c\sim n_s^\infty\sim (N/\log N)^{1/2}$.

\subsection{Static dynamics}
The static dynamics starts with the initial condition
that all nodes are occupied, $n_0=N$ ($\rho_0=1$). We consider the
scaling behavior of $\rho_s$, the particle density averaged over surviving
samples.

The rate equation~(\ref{rate_eq}) takes a different form
depending on the particle density.
When $\rho k_{\max} \gg  1$, it becomes $d\rho/dt =
-c\rho^{\theta-1}/2$, which yields
\begin{equation}\label{rho_t_1}
\rho_s (t) \sim t^{-1/(\theta-2)} .
\end{equation}
If the density becomes sufficiently small such that
$\rho k_{\max}\ll 1$, then the rate equation should be replaced by $d\rho/dt =
-g \rho^2/2$, which yields the solution
\begin{equation}\label{rho_t_2}
\rho_s (t)\sim (gt)^{-1} .
\end{equation}
The crossover between the two regimes takes place at time
\begin{equation}\label{z_*}
t_* \sim g^{(\theta-2)/(3-\theta)}\sim N^{{\bar z}_*} ~\mbox{with}~{\bar z}_* = (\theta-2)/w .
\end{equation}
At this crossover time scale $t=t_*$, the system starts to feel the finite upper bound
of the maximum degree, $k_{\max}$.
The density at the crossover is given by
\begin{equation}\label{alpha*}
\rho_* \sim g^{-1/(3-\theta)} \sim N^{-\alpha_*} ~\mbox{with}~ \alpha_* = 1/\omega .
\end{equation}

\begin{figure}[t]
\includegraphics[width=\columnwidth]{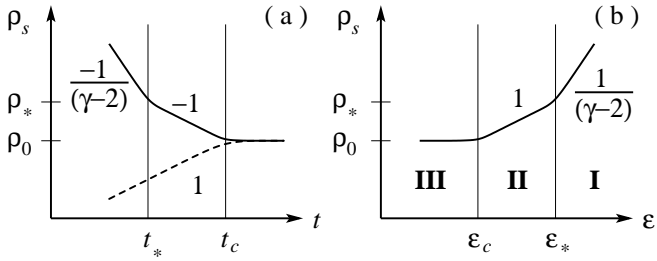}
\caption{(a) Schematic plot of $\rho_s$ vs. $t$ in the log-log scale at the
critical point in the annealed networks with $2<\gamma<3$ and
$\omega>\gamma-1$. The solid~(dashed) line corresponds
to the static~(defect) dynamics.
(b) Schematic plot of $\rho_s$ vs. $\varepsilon$ in the same condition.
}\label{fig1}
\end{figure}

Finally, the system reaches the stationary state.
From Eq.~(\ref{n_sc}), the particle density at the stationary state
is given by
\begin{equation}\label{alpha}
\rho_s^\infty \sim \sqrt{1/(gN)}\sim N^{-\alpha} ~\mbox{with}~
 \alpha = (1+(3-\theta)/\omega)/2.
\end{equation}
The saturation time $t_c$ determined from $(gt_c)^{-1} \sim N^{-\alpha}$
has the same scaling behavior as the relaxation time
in the defect dynamics (see in Eqs.~(\ref{defect_tc})and (\ref{z_d})). This means that the finite systems reach
the stationary state at the same time scale, irrespective of the initial
conditions.

There are a few remarks.
For $\omega > \gamma-1$ and $\gamma<3~(\theta=\gamma)$, there exist two distinct
$N$-dependent time scales $t_* \sim N^{{\bar z}_*}$ and $t_c\sim N^{\bar z}$ with ${\bar z}_*<{\bar z}$.
The former comes into play due to the finiteness of the maximum degree
$k_{\max}\sim N^{1/\omega}$, while the latter is the time scale to reach
the stationary state in finite networks.
This implies that finite-size effects in the annealed SF networks
depends on the limiting procedure how $N$ and $k_{\max}$ are taken to infinity.
For $\gamma>3~(\theta=3)$, the distinction  between the first regime
(Eq.~(\ref{rho_t_1})) and the second regime~(Eq.~(\ref{rho_t_2}))
disappears. The particle density decays as $\rho_s\sim t^{-1}$ for
$t<N^{\bar z}$ with ${\bar z}=1/2$, and then saturates to the stationary state value
$\rho_s^\infty \sim N^{-1/2}$. At $\gamma=3$, $\rho_s\sim (t\log t)^{-1}$ for $t<t_c\sim (N/\log N)^{1/2}$
and  $\rho_s^\infty\sim (N\log N)^{-1/2}$.

The systems with the natural cutoff ($\omega=\gamma-1$) are special.
Even for $\gamma<3$, the two time scales $t_*$ and $t_c$
coincide having ${\bar z}_*={\bar z} = (\gamma-2)/(\gamma-1)$.
This means that the second regime does not exist.
The density decays as $\rho_s \sim t^{-1/(\gamma-2)}$ for
$t<N^{\bar z}$, and then saturates to $\rho_s^\infty \sim N^{-1/(\gamma-1)}$.

The defect and static dynamics at criticality are illustrated
schematically in Fig.~\ref{fig1}(a).

\subsection{Numerical simulations}
We have performed numerical simulations in the annealed SF networks
to confirm the analytic results.
In the defect simulations, the survival probability is expected to scale as
\begin{equation}\label{p_scale}
P_s(t,N) = N^{-{\bar z} \delta} \mathcal{P}(t/N^{\bar z}) .
\end{equation}
The scaling function
behaves as $\mathcal{P}(x) \sim x^{-\delta }$ as $x\rightarrow 0$
and decays exponentially as $x\rightarrow \infty$.
The particle number averaged over surviving samples is expected to scale as
\begin{equation}\label{n_scale}
n_s(t,N) = N^{{\bar z} \tilde\eta } \mathcal{N}(t/N^{\bar z}) .
\end{equation}
The scaling function behaves as $\mathcal{N}(x)\sim x^{\tilde\eta}$ as $x\rightarrow 0$
and converges to a constant as $x\rightarrow \infty$.

\begin{figure}[t]
\includegraphics*[width=\columnwidth]{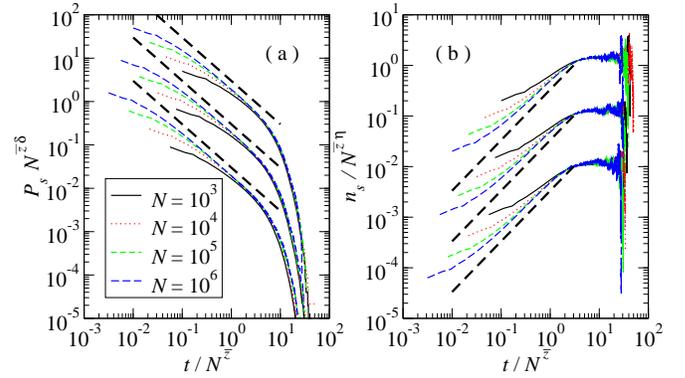}
\caption{(Color online) (a) Scaling plot for the survival probability. (b) Scaling plot
for the number of particles averaged over surviving samples.
The upper, middle, and lower sets of data correspond to the annealed
SF networks with $\gamma=2.5$ and $\omega=1.5$, $2.0$, and $3.0$,
respectively.
For readability, each data set is scaled down by a constant factor.
The dashed lines are guides to the eyes having slope $-1$ in (a)
and $1$ in (b).}\label{fig2}
\end{figure}

Figure~\ref{fig2} shows the scaling plots according to the scaling form in
Eqs.~(\ref{p_scale}) and (\ref{n_scale}) with the exponent values in
Eqs.~(\ref{delta}), (\ref{eta}), and (\ref{z_d}). The annealed networks of
size $N=10^3,\cdots,10^6$ were generated with the deterministic method.
The plotted data were obtained by averaging over $10^6$ runs. The nice data
collapse confirms the validity of the analytic result.

\begin{figure}[t]
\includegraphics*[width=\columnwidth]{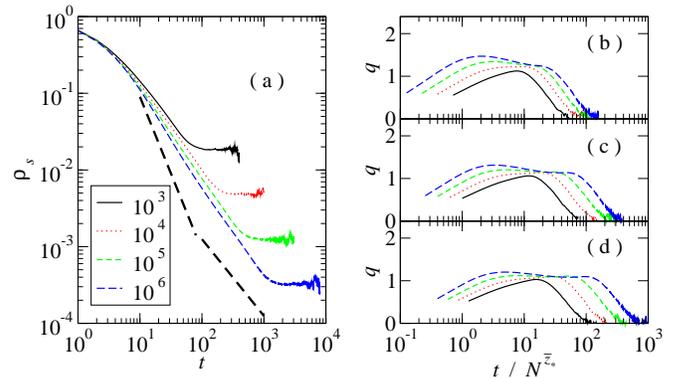}
\caption{(Color online) (a) Critical density decay at $\gamma=2.5$ and $\omega=2.5$. The
network sizes are $N=10^3,\cdots,10^6$. Effective exponent $q$ versus
$t/N^{{\bar z}_*}$ at $\gamma=2.5$ and (b)~$\omega=2.0$, (c)~$2.5$, (d)~$3.0$.}
\label{fig3}
\end{figure}

We have also performed the static simulations. Numerical data at the
critical point obtained in networks
with $\gamma=2.5$ and $\omega=2.5$ are presented
in Fig.~\ref{fig3}(a). Unlike in the schematic plot in Fig.~\ref{fig1}, the
crossover between two regimes with $\rho_s \sim t^{-1/(\gamma-2)}$ and
$\rho_s \sim t^{-1}$, respectively, is not prominent. Moreover, the decay
exponents seem to deviate from the expected values significantly. In order
to understand the origin of the discrepancy, we have performed a local slope
analysis. As an estimate for the density decay exponent, we define an
effective exponent $q(t) \equiv - \ln(\rho_s(t)/\rho_s(t/m)) / \ln m$
with a constant $m=4$.
The effective exponents measured at $\omega=2.0$, $2.5$, and $3.0$ are
plotted in Figs.~\ref{fig3}(b), (c), and (d), respectively, against
the scaling variable $ x=t/N^{{\bar z}_*}$. The analytic theory predicts that $q$
should converge to $1/(\gamma-2)=2$~($1$) as $N$ increases for
$x< 1$~($x> 1$). The effective exponent plot shows a weak but clear
tendency toward the analytic prediction. For $x<1$, the effective exponents steadily increases above 1 with network size, but still much lower than the predicted value 2 even for $N=10^6$. Moreover there is no appreciable power-law region (flat region for $q$). For $x>1$, the effective exponents overshoot the predicted value 1 till $N=10^5$, but start to decrease slightly at $N=10^6$. We also notice the appreciable flat region in this case.

In order to identify numerically the power-law scaling in the first regime, it would be required that at least $t_*\sim 10^2$ (two log decades). With $\gamma=2.5$ and $w=2.5$, the system size must be larger than $~\sim 10^{10}$, which is beyond the current computer capacity.

We have also studied the FSS behavior of the stationary-state particle density at criticality.
It exhibits a power-law scaling with $N$ as $\rho_s^\infty \sim N^{-\alpha}$.
It is found (not shown here) that the numerical result for the exponent $\alpha$ is compatible with
the analytic result given in Eq.~(\ref{alpha}). However, a discrepancy
becomes noticeable as $\gamma$ becomes smaller and $\omega$ becomes larger,
due to  strong finite size effects.
The degree distribution becomes singular as $\gamma$ approaches $2$.
At large $\omega$, the maximum degree $k_{\max}\sim N^{1/\omega}$ grows so
slowly that it becomes difficult to observe the asymptotic scaling behavior.

\section{Off-critical scaling}
For $2<\gamma<3$, the particle density exhibits distinct dynamic
characteristics depending on whether $\rho k_{\max}>1$ or $\rho k_{\max}<1$. This causes
an interesting cutoff-dependent FSS behavior at the critical
point. Such a cutoff dependence disappears far from the critical point.
However, in finite systems near the critical point, the
cutoff-dependence can still survive to lead to an anomalous FSS behavior.
For $\gamma>3$, the system shows a simple normal FSS behavior.

Near the critical point at $p=p_c(1-\varepsilon)$, the rate equation for the
density, Eq.~(\ref{rate_eq}), reads for $2<\gamma<3$
\begin{eqnarray}
d\rho_s/dt &= \varepsilon \rho_s - c' \rho_s^{\gamma-1} \ \ &\mbox{for}\
\rho_s k_{\max}>1, \label{OC1}\\
 &= \varepsilon \rho_s - \frac{1}{2} g \rho_s^2 \ \ &\mbox{for}\ \rho_s k_{\max}<1  ,\label{OC2}
\end{eqnarray}
where $k_{\max}=N^{1/\omega}$, $g=\langle k^2\rangle / \langle
k\rangle^2 \sim N^{(3-\gamma)/\omega}$,  and $c'=c/2$.

By setting $d\rho_s/dt = 0$, one obtains that the stationary
state solution is given by
\begin{eqnarray}
\rho_s^\infty &\sim  \varepsilon^{\beta} \ \ &\mbox{for} \ \ \rho_s k_{\max} >1
,\\
&\simeq  2\varepsilon / g \ \ &\mbox{for} \ \ \rho_s k_{\max} < 1\label{linear}
\end{eqnarray}
with the bulk order parameter exponent
\begin{equation}
\beta = 1/(\gamma-2) .
\end{equation}
This shows that the stationary state solution also depends on the degree
cutoff $k_{\max}$.
There is a crossover at $\varepsilon=\varepsilon_*$ with
\begin{equation}
\label{eq-eps1}
\varepsilon_* \sim
g^{-(\gamma-2)/(3-\gamma)} \sim N^{-1/\bar\nu_*} \mbox{with}~ 1/\bar\nu_* =
\frac{\gamma-2}{\omega} .
\end{equation}
For $\varepsilon < \varepsilon_*$, the order parameter scaling changes into the $\gamma$-independent ordinary MF linear scaling as in Eq.~(\ref{linear}), although this crossover disappears ($\varepsilon_*\rightarrow 0$) in the
thermodynamic limit.

When $\varepsilon$ decreases further below $\varepsilon_*$, the system will reach
the critical state where the particle density scales as
$\rho_s^\infty \sim  \sqrt {1/(gN)}$~(see Eq.~(\ref{alpha})).
The critical region in finite systems starts at $\varepsilon = \varepsilon_c$
with
\begin{equation}\label{eq-epsc}
\varepsilon_c\sim \sqrt{\frac{g}{N}} \sim N^{-1/\bar\nu} \mbox{with}~
1/\bar\nu= \frac{1-(3-\gamma)/\omega}{2},
\end{equation}
where the finite size saturation starts to occur
($2\varepsilon_c/g=\rho_s^\infty$).

The off-critical FSS behavior is illustrated in Fig.~\ref{fig1}(b).
This scaling theory predicts that there exist two characteristic sizes
$N_* \sim \varepsilon^{-\bar\nu_*}$ and $N_c \sim
\varepsilon^{-\bar\nu}$ which separate three scaling regimes.
In the regime I ($\varepsilon > \varepsilon_*$) where $N_* < N$,
the system behaves as in a SF network with infinite $N$ and infinite $k_{\max}$,
e.g.~$\rho_s\sim \varepsilon^\beta$.
In the regime II ($\varepsilon_c < \varepsilon < \varepsilon_*$)
where $N_c < N < N_*$, it behaves as in a SF network with infinite $N$
but with finite $k_{\max}$. The density scales as $\rho_s \simeq 2\varepsilon /
g \sim \varepsilon N^{-(3-\gamma)/\omega} $.
Finally, it behaves as in a SF network with finite $N$ and $k_{\max}$ in
the regime III ($\varepsilon<\varepsilon_c$) where $N < N_c$.
The density scales as $\rho_s \sim N^{-\alpha}$ with $\alpha=(1+(3-\gamma)/\omega)/2$
(see Eq.~(\ref{alpha})).

At the special case of the natural cutoff with $\omega=\gamma-1$, the regime II disappears
and there is a direct crossover from the regime I (no size effect) to the regime III (critical size scaling).
For $\gamma>3$ where $\theta=3$ and $g\sim O(1)$, $\rho_s^\infty \sim \varepsilon$ in both the regime I and II, so $\varepsilon_*$ becomes meaningless. Here again we observe a direct crossover from the regime I to the regime III.

The FSS scaling behavior in the annealed SF network is sharply contrasted
with that in the quenched SF network. While there are two
characteristic sizes $N_*$ and $N_c$ that depend explicitly on the
degree cutoff in the former (at least for $2<\gamma<3$ and $\omega>\gamma-1$), it has been proposed in the latter through
a droplet-excitation (hyperscaling) argument~\cite{Hong07}
that there exists a unique cutoff-independent characteristic size $N_q
\sim \varepsilon^{-\bar\nu_q}$ with $1/\bar\nu_q = (\gamma-2)/(\gamma-1)$ for $2<\gamma<3$
and $1/\bar\nu_q=1/2$ for $\gamma>3$. It is interesting to notice that
the FSS theory in the annealed network
coincides with that in the quenched network for $\gamma>3$ and also at the special case with
the natural cutoff with $\omega=\gamma-1$ for $2<\gamma<3$.
The origin of the discrepancy in the FSS theory between two different networks
as well as the relevance/role of the quenched linking disorder have not been fully
explored as yet, which awaits a further investigation.

\begin{figure}
\includegraphics*[width=\columnwidth]{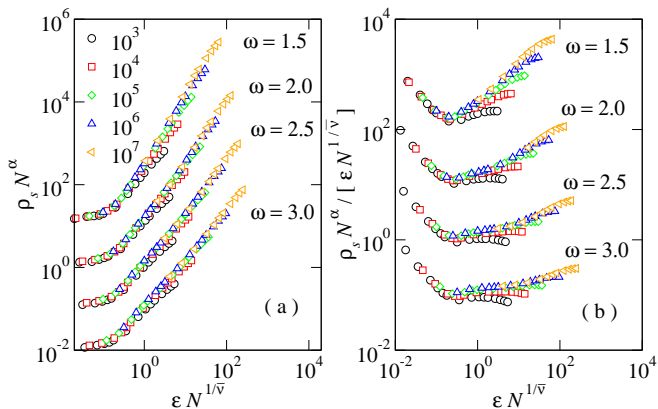}
\caption{(Color online) (a) Scaling plots of $y = \rho_s N^\alpha$ vs.
$x =\varepsilon N^{1/\bar\nu}$
at $\gamma=2.5$ and $\omega = 1.5, 2.0, 2.5$, and $3.0$. Network sizes are
$N=10^3,\cdots,10^7$.
(b) Plots of $y/x$ against $x$.  Each data set is shifted
vertically by a constant factor to avoid an overlap.}\label{fig4}
\end{figure}

We have performed extensive simulations in the annealed SF networks to test
the off-critical FSS theory.
In Fig.~\ref{fig4}, we present a scaling plot of $y \equiv \rho_s N^\alpha$
against a scaling variable $x\equiv \varepsilon N^{1/\bar\nu}$ at $\gamma=2.5$
and $\omega=1.5, 2.0, 2.5$, and $3.0$.
When $\omega=\gamma-1=1.5$ (natural cutoff), the FSS theory predicts that the quantity $y$
converges to a constant value for $x\ll 1$~(regime III) and scales as $y\sim
x^{1/(\gamma-2)}=x^2$ for $x\gg 1$~(regime I). There should be no regime II.
The numerical data in
Fig.~\ref{fig4} seem to support this two-regime scaling behavior reasonably well.

When $\omega>\gamma-1$, we expect three scaling regimes. Numerical data in
the regime II and III will converge to a single curve, but those in the
regime I should deviate from it because of the two different
characteristic sizes. The numerical data
in Fig.~\ref{fig4}(a) show a clear evidence of the regime III (flat region), but a weak
signature of the regimes II (linear-slope region, $y\sim x$) and I (no collapse). Although the signature is not prominent due to strong finite size effects,
the existence of the three scaling regimes is evident. In Fig.~\ref{fig4}(b),
we present the numerical data in a different style by plotting $y/x$ against
$x$, so the regime II can be identified by a flat region. As expected, there is no
flat region at $\omega=1.5$ (natural cutoff). As $\omega$ increases,
one can see clearly broadening of the flat region (regime II) which becomes larger with increasing $N$.
This behavior is qualitatively consistent with the expected FSS behavior.
It is very difficult to observe the regime III scaling even at $N=10^7$, similar to the difficulty encountered in the study of the critical dynamics in Sec.~\ref{CD}.

Finally, the off-critical dynamic behavior can be easily derived from the rate equations, (\ref{OC1}) and (\ref{OC2}), in the thermodynamic limit, approaching the criticality from the active or the absorbing side:
$\rho_s(t)-\rho_s^\infty\sim e^{-t/\tau}$ where the relaxation time scales as
$\tau\sim \varepsilon^{-\nu_t}$ with $\nu_t=1$ in all cases. These results are consistent with
our previous results through relaxation time relations of ${\bar z}=\nu_t/\bar\nu$ and ${\bar z}_*=\nu_t/\bar\nu_*$.

\section{Sample-to-sample fluctuations}
Suppose that one wants to generate a network of $N$ nodes with a given
degree distribution $P(k)$ with a (forced or natural) cutoff. In general, there are two kinds of quenched
disorder to be considered. First, one should sample a degree sequence $\{k_1,\cdots,k_N\}$ from $P(k)$
and then choose a way of linking the nodes together to create a network.
Disorder can be involved in both processes, which is named as sampling disorder and  linking disorder, respectively.
A {\em quenched} network involves both sampling and linking disorder, in general.

 An annealed network is free
from the linking disorder. However, it may still have the sampling disorder.
In the numerical studies in the preceding sections, we have sampled the
degree sequence deterministically without any disorder. On the other hand,
probabilistic sampling of the degree sequence leads to the sampling
disorder. In this section, we investigate sample-to-sample fluctuations
in annealed networks due to the sampling disorder.
The quantity of our primary interest is $g\equiv \langle k^2 \rangle / \langle
k\rangle^2$.

When $N$ values $\{k_1,\cdots,k_N\}$ are drawn probabilistically in accordance with the distribution
$P(k)$ for $k_{\min} \le k \le k_{\max}$, a sampled distribution
$\tilde P(k)=\sum_{i=1}^N \delta_{k,k_i}/N$ may
deviate from the target distribution $P(k)$ due to the finiteness of $N$. The deviation is denoted
by $\delta P(k) = \tilde  P(k) - P(k)$. Then, it is straightforward
to show that
\begin{eqnarray}
\left[\delta P(k) \right] &=& 0  \label{correlator1} \\
\left[\delta P(k) \delta P(k') \right] &=&
- \frac{P(k)P(k')}{N} + \frac{P(k)}{N} \delta_{k,k'} , \label{correlator2}
\end{eqnarray}
where $[\cdots]$ denotes the sample (disorder) average.

The $n$th moment of the degree of a sample is given by
\begin{equation}
\langle k^n \rangle \equiv \sum_k k^n \tilde P(k) = \langle n \rangle_0 \left( 1
+ \frac{\langle n\rangle_\delta}{\langle n\rangle_0} \right),
\end{equation}
where we introduce shorthand notations as
$\langle n\rangle_0 \equiv \sum_{k}k^n P(k)$ and $\langle
n\rangle_\delta \equiv \sum_{k}k^n \delta P(k)$.
There is an $1/N$ factor in the correlator in Eq.~(\ref{correlator2}).
So, $\langle n\rangle_\delta/\langle n\rangle_0$ can be considered as a small
expansion parameter for large $N$. Up to the second order, the quantity $g$ of a given
sample can be written as
$$
g = \frac{\langle 2\rangle_0}{\langle 1\rangle_0^2} \left(
1+\frac{ \langle 2\rangle_\delta}{\langle 2 \rangle_0}
- 2 \frac{\langle 1\rangle_\delta}{\langle 1\rangle_0}
+ 3 \frac{\langle 1\rangle_\delta^2}{\langle 1\rangle_0^2}
- 2 \frac{\langle 2\rangle_\delta \langle 1\rangle_\delta}{\langle
  2\rangle_0 \langle 1\rangle_0} \right) .
$$
The disorder-averaged correlators in Eqs.~(\ref{correlator1}) and (\ref{correlator2})
imply that $\left[ \langle n\rangle_\delta \right]=0$ and that
\begin{equation}
\left[ \langle m\rangle_\delta \langle n \rangle_\delta \right]
= \frac{1}{N} \left( \langle m+n \rangle_0 - \langle m\rangle_0 \langle
n\rangle_0 \right) .
\end{equation}
This allows us to expand systematically $[g]$ and $(\Delta g)^2 \equiv [g^2] -
[g]^2$ in powers of $\frac{1}{N}$. After some algebra, we obtain the
following result up to the order $1/N$:
\begin{equation}
[g] = \frac{\langle 2\rangle_0}{\langle 1\rangle_0^2} \left\{
1 + \frac{1}{N} \left( 3 \frac{\langle 2\rangle_0}{\langle 1\rangle_0}
- 2 \frac{\langle 3 \rangle_0}{\langle 1\rangle_0 \langle 2\rangle_0} - 1
\right) \right\}
\label{g_1}
\end{equation}
and
\begin{equation}
(\Delta g)^2 = \frac{1}{N} \frac{\langle 2\rangle_0^2}{\langle 1\rangle_0^4}
\left\{ \frac{\langle 4\rangle_0}{\langle 2\rangle_0^2}
- 4 \frac{\langle 3\rangle_0}{\langle 1\rangle_0 \langle 2 \rangle_0}
+ 4 \frac{\langle 2\rangle_0}{\langle 1\rangle_0^2} -1 \right\} .
\label{g_2}
\end{equation}

Our interest lies in the SF network of $N$ nodes having the degree
distribution $P(k) \propto k^{-\gamma}$ in the interval $k_{\min}\le k\le
k_{\max}=N^{1/\omega}$ with $\gamma>2$ and $\omega\ge \gamma-1$.
The $1/N$ term in Eq.~(\ref{g_1}) is always subleading, so the scaling behavior of $[g]$
is determined by $\langle 2 \rangle_0$, which yields that
\begin{equation}
[g] = \left\{
\begin{array}{lll}
\sim N^{(3-\gamma)/\omega} & \mbox{for}& 2<\gamma<3 \\ [2mm]
\sim \log N & \mbox{for}& \gamma=3 \\ [2mm]
\sim O(1) &\mbox{for}& \gamma >3 .
\end{array} \right.
\end{equation}
On the other hand, the term $\langle 4\rangle_0 / \langle
2\rangle_0^2$ in the parenthesis of Eq.~(\ref{g_2}) makes a leading order
contribution. Hence, we find that the relative variance
$R_g = (\Delta g)^2 / [g]^2$ is given by
\begin{equation}\label{d_f}
R_g = \left\{
\begin{array}{lll}
\sim N^{ (\gamma-1)/w - 1 } & \mbox{for} & 2<\gamma < 3 \\ [2mm]
\sim N^{ 2/w - 1 }(\log N)^{-2} & \mbox{for} & \gamma = 3 \\ [2mm]
\sim N^{ (5-\gamma)/w -1 } & \mbox{for} & 3 < \gamma < 5 \\
[2mm]
\sim N^{-1} \log N &\mbox{for}&  \gamma =5 \\ [2mm]
\sim N^{-1}  &\mbox{for}&  \gamma >5 .
\end{array} \right.
\end{equation}

In the theory of disordered systems, the relative variance $R_X$ of an
observable $X$ due to a quenched disorder is an indicator of the
self-averaging property~\cite{Aharony96}. When it vanishes in the thermodynamic limit
$N\rightarrow\infty$, such a system is said to be {\em self-averaging}.
The self-averaging property implies that an observable measured
in a sample with a typical disorder configuration takes the same value as the
sample-averaged value in the $N\rightarrow\infty$ limit.
A system with $R_X \sim N^{-1}$ is said to be
{\em strongly self-averaging}~(SSA). This is the case when the central limit
theorem works. When $R_X \sim N^{-r}$ with $r<1$, such a system is
said to be {\em weakly self-averaging}~(WSA).
A system with strong or relevant disorder lacks the self-averaging property near
the criticality.
In such a system, $R_X$ converges to a finite value as $N$ increases.

The result in Eq.~(\ref{d_f}) discloses the self-averaging property of the
annealed SF network under the sampling disorder. First of all, we find that
the system with $\gamma>5$ is SSA at all values of the degree cutoff
exponent $\omega$. For $\gamma\le 5$, $R_g$ decays slower than $N^{-1}$
at all values of $\omega>\gamma-1$. So the system is WSA.

Interestingly, the systems lack the self-averaging property
when $2< \gamma<3$ and $\omega=\gamma-1$ ($R_g$ approaches a non-zero constant as $N$ increases).
Note that the cutoff exponent
$\omega=\gamma-1$ corresponds to the natural cutoff. Networks without
explicit constraint on the degree also display the cutoff scaling
$k_{\max} \sim N^{1/(\gamma-1)}$.
In these networks, not only the node-to-node degree
fluctuation but also the sample-to-sample degree fluctuations are very
strong.

\begin{figure}
\includegraphics*[width=\columnwidth]{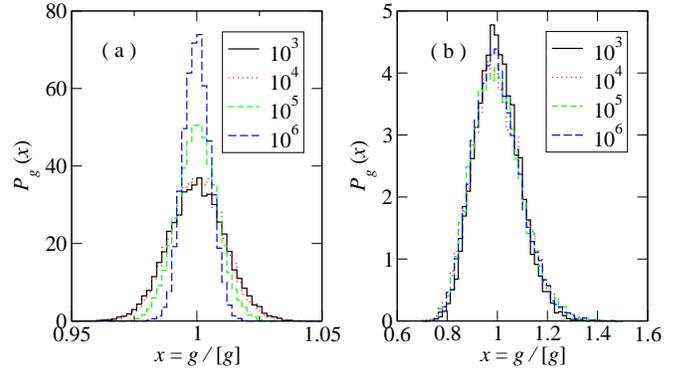}
\caption{(Color online) Probability distribution $P_g(x)$ for $x=g/[g]$ in the annealed
networks with $N=10^3,\cdots,10^6$. (a) $\gamma=2.75$ and $\omega=3.0$ (b)
$\gamma=2.75$ and $\omega=1.75$.}\label{fig5}
\end{figure}

We present numerical data showing the (non-)self-averaging property in
Fig.~\ref{fig5}. Drawing $N$ values of $k$ from the distribution $P(k)\sim
k^{-\gamma}$ in the interval $2\le k\le N^{\omega}$, we calculated
$g=\langle k^2\rangle / \langle k\rangle^2$. This was repeated
$N_S=10^5$ times, from which one can construct the probability distribution
function $P_g(x)$ for $x=g/[g]$. Figure~\ref{fig5}(a) shows that the
distribution becomes sharper and sharper as $N$ increases. It indicates the
self-averaging property at $\gamma=2.75$ and $\omega=3.0$. On the other
hand, Fig.~\ref{fig5}(b) shows that the distribution converges to a
limiting distribution. It indicates that the system is not self-averaging
at $\gamma=2.75$ and $\omega=\gamma-1=1.75$.

The strong disorder fluctuation raises an important question.
In general, a real complex network is a disordered media having quenched
disorder, for example, the sampling disorder and the linking disorder as
mentioned before.
Being coupled with dynamic degrees of freedom, the quenched structural
disorder may give rise to disorder-relevant critical phenomena.
This is a plausible scenario, but has been ignored in most studies.
It seems a quite challenging problem to incorporate the quenched disorder
into a systematic  analysis.

In the annealed network considered in this study, the dynamic degrees of
freedom is completely decoupled with the sampling disorder (no linking disorder). Hence, the
scaling theory developed here should be valid whether the sampling
disorder is self-averaging or not. However, our result still warns that
the sample average of any observable involving $g$
is practically meaningless due to its broad distribution, which occurs in the critical region
(regime II and III) in annealed networks with $2< \gamma<3$ and the natural cutoff.

\section{Discussion and Summary}

We studied critical behavior of the CP in annealed scale-free networks.
For the degree exponent $\gamma>3$, the standard single-parameter FSS is found with
various dynamic and static exponents which are independent of the cutoff exponent $\omega$
and also $\gamma$. For highly heterogeneous networks with $\gamma<3$, there exist two
different characteristic time scales and their associated exponents depend not only on $\gamma$
but also on $\omega$. These results are contrasted with those in quenched scale-free networks
where a single-parameter FSS is found without any cutoff dependence even for $\gamma<3$ if the cutoff is not strong enough ($\omega<\gamma$)~\cite{Hong07,Ha07}. At the special case of $\omega=\gamma-1$ (natural cutoff), these two different FSS  coincide to each other.

Annealed networks may include the sampling disorder, which generates a strong sample-to-sample
fluctuation in highly heterogeneous networks with the natural cutoff. In quenched networks, the linking disorder is inherent, which generates the density-density correlation in neighboring nodes through coupling with fluctuating variables. This correlation leads to the shift of the transition point of the CP model~\cite{CPS06,Ha07}. In addition, the linking disorder generates another type of sample-to-sample fluctuations
which cause spreading of the transition points in finite systems. Hong, Ha, and Park~\cite{Hong07} showed that there exists a characteristic (droplet) size scale diverging as $N_q\sim \varepsilon^{-\bar\nu_q}$ with $1/\bar\nu_q=(\gamma-2)/(\gamma-1)$ for $\gamma<3$. For $N<N_q$ (or equivalently $\varepsilon < \varepsilon_q$ with $\varepsilon_q\sim N^{-(\gamma-2)/(\gamma-1)}$), the system feels the droplet length scale and the finite-size effect is dominant. As $\varepsilon_q > \varepsilon_c$ given in Eq.~(\ref{eq-epsc}), one may expect that the finite-size saturation induced by the droplet length scale comes in earlier (at $\varepsilon=\varepsilon_q$) in quenched networks than in annealed networks. Then, the cutoff-dependency of the saturation density may disappear. However, as $\varepsilon_q < \varepsilon_*$ given in Eq.~(\ref{eq-eps1}), the cutoff-dependent density-decaying dynamics comes in before saturation. The linking disorder fluctuation may be responsible for the disappearance of this dynamics in the quenched networks, but this is just a speculation as yet. A full understanding for the FSS behavior in quenched networks needs a further investigation.

During the final stage of preparing this manuscript, Bogu\~n\'a, Castellano, and Pastor-Satorras posted a preprint on the cond-mat archive~\cite{BCPS08}, the results of which partially overlap with those presented here.

\acknowledgments
This work was supported by KOSEF grant Acceleration Research (CNRC)
(Grant No. R17-2007-073-01001-0). This work was also supported by the Korea
Research Foundation grant funded by MEST (Grant No. KRF-2006-003-C00122).


\begin{thebibliography}{50}
\bibitem{Dorogovtsev07} S.N. Dorogovtsev, A.V. Goltsev, and J.F.F. Mendes,
        arXiv:0705:0010 (2007).
\bibitem{WS98} D.J. Watts and S.H. Strogatz,
        Nature {\bf 393}, 440 (1998).
\bibitem{BA99} A.-L. Barab\'asi and R. Albert
        Science {\bf 286}, 509 (1999).
\bibitem{Hong07} H. Hong, M. Ha, and H. Park,
        Phys. Rev. Lett. {\bf 98}, 258701 (2007).
\bibitem{Castellano08} C. Castellano and R. Pastor-Satorras,
        Phys. Rev. Lett. {\bf 100}, 148701 (2008).
\bibitem{Hin00} H. Hinrichsen, Adv.Phys. {\bf 49}, 815 (2000).
\bibitem{Goh01} K.-I. Goh, B. Kahng, and D. Kim,
        Phys. Rev. Lett. {\bf 87}, 278701 (2001).
\bibitem{Catanzaro05} M. Catanzaro, M. Bogu\~n\'a, and R. Pastor-Satorras,
        Phys. Rev. E {\bf 71}, 027103 (2005).
\bibitem{HKPP98} W.M. Hwang, S. Kwon, H. Park, and H. Park, Phys. Rev. E {\bf 57}, 6438 (1998).
\bibitem{Fisher84} M.E. Fisher, J. Stat. Phys. {\bf 34}, 667 (1984).
\bibitem{Aharony96} A. Aharony and A.B. Harris,
        Phys. Rev. Lett. {\bf 77}, 3700 (1996).
\bibitem{Ha07}  M. Ha, H. Hong, and H. Park,
        Phys. Rev. Lett. {\bf 98}, 029801 (2007).
\bibitem{CPS06} C. Castellano and R. Pastor-Satorras,
        Phys. Rev. Lett. {\bf 96}, 038701 (2006).
\bibitem{BCPS08} M. Bogu\~n\'a, C. Castellano, and R. Pastor-Satorras,
        Phys. Rev. E {\bf 79}, 036110 (2009) (e-print arXiv:0810.3000).
\end{thebibliography}
\end{document}